\documentclass[aps,prb,twocolumn,superscriptaddress,showpacs]{revtex4-2}
\usepackage{amsmath,amssymb,amsfonts,amsthm}
\usepackage{braket}
\usepackage{graphicx,import}
\usepackage{bm}
\usepackage{bbm}
\usepackage{color}
\usepackage{booktabs,tabularx}
\usepackage{dcolumn}   
\usepackage{epstopdf}
\usepackage{bbold}
\usepackage[colorlinks=true,linkcolor=blue,urlcolor=blue,citecolor=blue]{hyperref}
\usepackage{commath}
\usepackage[caption=false]{subfig}
\usepackage{amsmath}
\usepackage{soul}

\begin{document}

\preprint{APS/123-QED}

\title{Spatiotemporal Quenches for Efficient Critical Ground State Preparation in Two-Dimensional Quantum Systems}

\author{Simon Bernier}
\email{simon.bernier@mail.mcgill.ca}
\affiliation{Department of Physics, McGill University, Montr\'eal, Qu\'ebec H3A 2T8, Canada}
\author{Kartiek Agarwal}%
\email{kagarwal@anl.gov}
\affiliation{Material Science Division, Argonne National Laboratory, Lemont, Illinois 60439, USA}
\affiliation{Department of Physics, McGill University, Montr\'eal, Qu\'ebec H3A 2T8, Canada}

\date{\today}

\begin{abstract}
Quantum simulators have the potential to shed light on the study of quantum many-body systems and materials, offering unique insights into various quantum phenomena. While adiabatic evolution has been conventionally employed for state preparation, it faces challenges when the system evolves too quickly or the coherence time is limited. In such cases, shortcuts to adiabaticity, such as spatiotemporal quenches, provide a promising alternative. This paper numerically investigates the application of spatiotemporal quenches in the two-dimensional transverse field Ising model with ferromagnetic interactions, focusing on the emergence of the ground state and its correlation properties at criticality when the gap vanishes. We demonstrate the effectiveness of these quenches in rapidly preparing ground states in critical systems. Our simulations reveal the existence of an optimal quench front velocity at the emergent speed of light, leading to minimal excitation energy density and correlation lengths of the order of finite system sizes we can simulate. These findings emphasize the potential of spatiotemporal quenches for efficient ground state preparation in quantum systems, with implications for the exploration of strongly correlated phases and programmable quantum computing.
\end{abstract}

\maketitle

\section{Introduction}
\label{sec:intro}

Modern quantum simulators hold immense potential for studying fundamental aspects of quantum many-body systems and materials. Recent experiments in ultracold atoms~\cite{bloch2008, bloch2012, gross2017, browaeys2020}, trapped ions~\cite{blatt2012,garttner2017,monroe2021}, photonic systems~\cite{Aspuru2012,Wang2020,Uppu2021} and superconducting qubits~\cite{Kjaergaard2020,Zhang2023}, among other promising platforms~\cite{Ladd2010,Altman2021,Weskesser2021,Kaufman2021} have successfully demonstrated many novel quantum phenomena. These include a variety of spin models~\cite{Georgescu2014, Jurcevic2015,keesling2019,jepsen2021,scholl2021,ebadi2021,scholl2022, Ma2014}, topological quantum numbers~\cite{tarruell2012,grusdt2013,Wang2019,leseleuc2019,wintersperger2020,semeghini2021}, many-body localization~\cite{pal2010,alet2018,schreiber2015,choi2016,smith2016,morong2021}, lattice gauge theories~\cite{zohar2016,haukeQS2013,martinez2016,banuls2020}, among others~\cite{choi2019,Tiarks2019,bluvstein2021,viermann2022,Feng2023,Enomoto2023}. A recent body of work has focused on simulating properties of two-dimensional systems \cite{Britton2012, Schreiber2012, Mielenz2016, Tang2018, Periwal2021, Rajabi2019}. These platforms have also emerged as candidates for programmable quantum computing~\cite{kasper2021,ebadi2022,Flamini2019,Politi2009,Defienne2016,Lu2019,Arrazola2019,Katz2023,Bluvstein2023}. A key application of such artificial quantum matter is to simulate strongly correlated phases of electrons in conventional materials~\cite{mazurenko2017,linke2018,tarruel2018,bohrdt2021}. Although Hamiltonians of many such systems can be approximately realized using a combination of fixed potentials and driving, it remains a challenge to prepare the system in a state corresponding to a low enough effective temperature at which the ground state properties can be reliably explored~\cite{mazurenko2017}.

Traditionally, the state preparation process involves adiabatic evolution~\cite{albash2018}, where the system is initially set in the ground state or a close approximation of an easily prepared Hamiltonian. By tuning the Hamiltonian parameters gradually, the state evolves towards the desired target state, often the ground state of a target Hamiltonian~\cite{albash2018}. If the evolution is slow enough, the quantum state remains in the ground state throughout. However, adiabatic preparation time scales with the square of the inverse of the smallest energy gap encountered during the tuning process. Consequently, when the gap closes, excitations are inevitably produced, and adiabatic techniques fail to reliably generate the target state.

In situations where adiabatic evolution is unsuccessful or too slow compared to the coherence time of the quantum simulator, shortcuts to adiabaticity become necessary. One such method is counter-diabatic driving, which employs auxiliary time-dependent Hamiltonians to counteract excitation production~\cite{delcampo2012,damski2014,sels2017}. Optimal control protocols, including bang-bang protocols, have been developed and rely on classical optimization of the protocol~\cite{pichler2018,ho2019ultra,pagano2020,ebadi2022}. Another approach involves spatially inhomogeneous quenches, where certain regions of the system act as sinks for excitations~\cite{ho2009,zaletel2021}.

For systems exhibiting emergent Lorentz symmetry, an efficient strategy for preparing the ground state of Hamiltonians is through spatiotemporal quenches~\cite{dziarmaga2010,agarwal2017,agarwal2018,mitra2019,sinha2020,bernier2022spatiotemporal}. These protocols enable the rapid production of ground states, even in critical cases characterized by linearly dispersing modes and vanishing energy gaps. In this protocol, the system starts in a low-entanglement state corresponding to the ground state of a Hamiltonian with a gapping perturbation. The perturbation is then tuned to zero along a quench front that moves at a time-dependent velocity greater than the speed of light. The simplest version of this protocol involves a constant velocity that optimally cools the system as it approaches the speed of light. Similar velocity thresholds have also been observed in systems with mobile defects~\cite{KarzigBoosting,Bastianello2018}. This approach has potential applications in simulating low-energy states of the Hubbard model in two dimensions~\cite{mazurenko2017,auerbach2012} and one-dimensional quantum gases described by a low-energy Luttinger liquid~\cite{giamarchi2003}.

\begin{figure}
    \includegraphics[width=\linewidth]{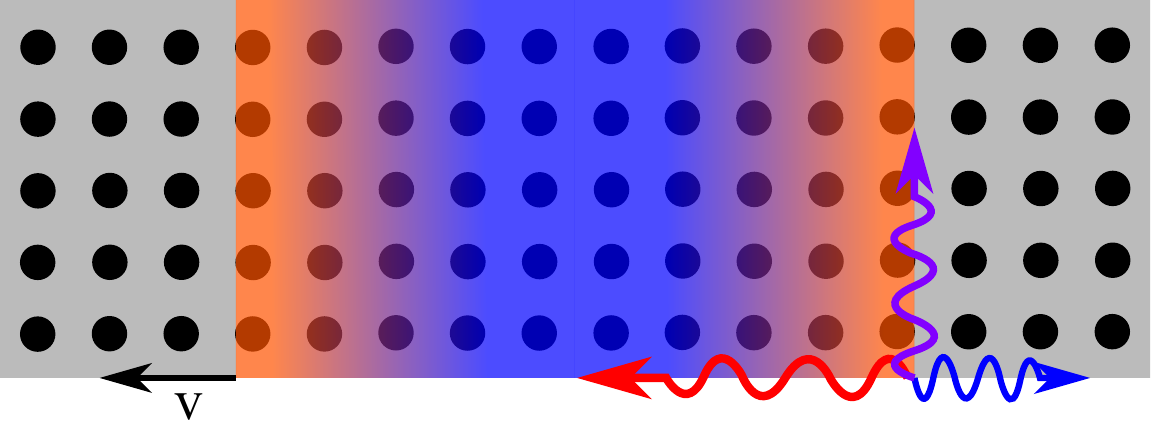}
    \caption{The quench front acts as a source of excitations, populating modes in a chiral way. At the end of the quench, the modes left in the wake of the front are populated according to Doppler shifted temperatures. Modes counter-propagating with or transverse to the front are populated at cold red-shifted temperatures, while modes copropagating with the front are populated according to a hot blue-shifted temperature.}
    \label{fig:2Dconcept}
\end{figure}

Intuitively, the spatiotemporal quench protocol exploits Doppler shifts to induce cooling. The quench front excites modes in a chiral manner, with co-propagating modes experiencing blue shifts and counter-propagating modes undergoing red shifts, as seen in Fig.~\ref{fig:2Dconcept}. As the velocity of the front approaches the speed of light, counter-propagating excitations are completely suppressed, and all energy is carried by excitations propagating with the quench front. Consequently, the resulting system exhibits critical ground state correlations. In two dimensions, even the transverse modes experience Doppler shifts, leading to the suppression of excitations, a fully relativistic effect. This method enables the preparation of ground states in critical models within a time that scales linearly with the system size, providing a significant advantage over adiabatic evolution, which requires a quadratic time scaling~\cite{agarwal2017}.

In this paper, we investigate spatiotemporal quenches in a two-dimensional system, focusing specifically on the ferromagnetic interactions of the two-dimensional transverse field Ising (2D-TFI) model. Previous theoretical studies have explored similar quenches in short-range and long-range TFI and Heisenberg models~\cite{dziarmaga2010,agarwal2018, bernier2022spatiotemporal}, which exhibit Lorentz invariance and a maximal speed of information propagation, as well as a linear causal light-cone. Experimental studies have mainly focused on anti-ferromagnetic interactions~\cite{Guardado-Sanchez2018,ebadi2021}. Our simulations demonstrate that the ``Doppler-shift" cooling persists when the ultraviolet modes remain unexcited by the quench front. Furthermore, we observe a local minimum in the energy density of excitations as a function of the quench front velocity, occurring around the speed of light. In this instance, the correlation length saturates with the system size, as confirmed by scaling collapse. These findings emphasize the effectiveness of spatiotemporal quenches and highlight that optimal cooling occurs when the quench front velocity approaches the emergent speed of light.

This manuscript is organized as follows. In Sec.~\ref{sec:model}, we introduce the model studied and its critical properties. In Sec.~\ref{sec:dopplerCooling}, we show that the energy density and correlations at the end of the quench are qualitatively consistent with the heatwave picture and Doppler cooling. We argue that the system reaches criticality at the emergent speed of light by doing an appropriate scaling collapse of the energy density and spin correlations. We discuss the growth of entanglement entropy during the quench, and conclude with a summary of findings and potential future directions in Sec.~\ref{sec:conclusion}. 

\section{Model studied}
\label{sec:model}

We study spatiotemporal quenches in the 2D-TFI model of $N=L_x \times L_y$ spins with ferromagnetic interactions,  $y$-periodic boundary conditions and Hamiltonian
\begin{equation} \label{eq:Hamiltonian}
    H = -J\left(\sum_{<i,j>} \sigma_i^x \sigma_j^x + g_c \sum_i \sigma_i^z \right) - h\sum_i f_i(t) \sigma_i^z,
\end{equation}
where $<i,j>$ refers to nearest-neighbours, $\sigma_i^\mu$ are the Pauli matrices, $J$ is the interaction strength, $g_c$ is the critical transverse field and $h$ is the initial gapping perturbation. In what follows, we set $J=1$, let $h=5g_c$ and $L_x = 8L_y$. The perturbation is quenched along smooth fronts moving at velocity $v$ such that $f_i(t) = \frac{1}{2} + \frac{1}{2}\tanh\!\left[(\abs{x_i}-vt)/v\tau\right]$, where $\tau$ is the smoothing parameter. The quench is started at time $t_0=-2\tau$ ensuring that $f_i(t_0)\approx 1$ at every site. At $t \rightarrow \infty$, $f_i (t) \rightarrow 0 \forall i$, resulting in the critical Hamiltonian. The quench time is halved by starting the quench in the center of the chain. In this paper, we restrict the study of quenches in systems up to $L_y=5$, giving $N=200$ spins.

The system is initialized in the ground state of the paramagnetic phase with large $h$. The initial wavefunction is obtained using ITensor's density matrix renormalization group (DMRG) algorithm~\cite{itensor}. The Hamiltonian is represented exactly as a matrix product operator (MPO). Next, time evolution is carried out with the fourth-order time-dependent variational principle~\cite{haegeman2011,haegeman2016}. At every time step, the spin correlations, the (von Neumann) entanglement entropy and the total energy are calculated using standard matrix product state (MPS) techniques~\cite{schollwock2011}. All quantities presented in this paper were found to converge for a maximum MPS bond dimension of $\chi=512$.

In this paper, we calculate the excitation energy density $\epsilon(t) = \frac{1}{N}(\langle H(t)\rangle-E_0(t))$ where $E_0(t)$ is the instantaneous ground state energy. We similarly compute the spatial distribution of the energy. The local energy density is defined over each bond between sites $(i,j), (i+1,j)$, as the expectation value of the operator
\begin{align}
    h_{i,j}(t) = &- \sigma^x_{i,j}\sigma^x_{i+1,j} - \frac{1}{2} \left( g_{i,j}\sigma^z_{i,j} + g_{i+1,j}\sigma^z_{i+1,j} \right) \nonumber\\
    &-\frac{1}{4} \left( \sigma^x_{i,j}\sigma^x_{i,j+1} + \sigma^x_{i+1,j}\sigma^x_{i+1,j+1} \right. \nonumber\\
    &\left. + \sigma^x_{i,j-1}\sigma^x_{i,j} + \sigma^x_{i+1,j-1}\sigma^x_{i+1,j} \right)
\end{align}
where $i\in [1,L_x-1]$, $j\in [1,L_y]$ and $g_{i,j} = g_c(1+hf_{i,j}(t))$, which corresponds to the instantaneous transverse field. The final term in this expression is a two dimensional interpolation of the energy contribution of the vertical bonds. In this manner, the spatial distribution of the excitation energy is defined as $\epsilon(x_i,y_j,t) = \langle h_{i,j}(t) \rangle - \epsilon_0(x_i,y_j,t)$, where $\epsilon_0(x_i,y_j,t)$ is the instantaneous ground state energy density. Note that at the boundaries, terms $\frac{1}{2}g_{i,j}\sigma^z_{i,j}$ must be added to $h_{i,j}$ for $i={1,N-1}$ such that $\sum_{i,j} h_{i,j} = H$. Thus, the local energy density sums up to the total energy.

\begin{figure}
    \centering
    \includegraphics[width=\linewidth]{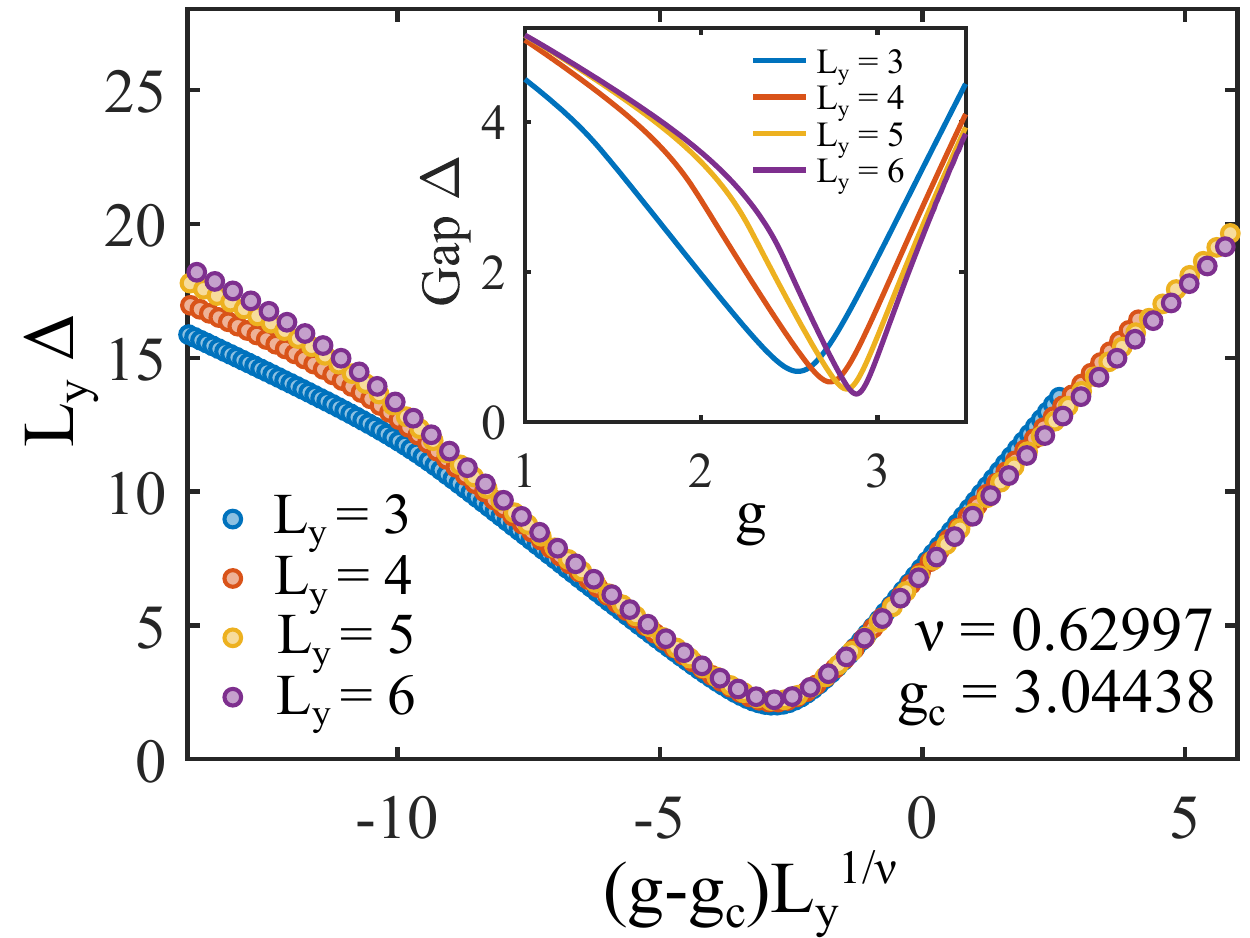}
    \caption{The critical properties of the 2D-TFI model are verified by doing a scaling collapse of the energy density near the minimum gap. The collapse is obtained for the known critical field $g_c = 3.04438$, critical exponent $\nu = 0.629971$ and dynamical critical exponent $z=1$. \textbf{Inset} The uncollapsed gap energy is shown.}
    \label{fig:energyGap}
\end{figure}

\begin{figure}
    \centering
    \includegraphics[width=\linewidth]{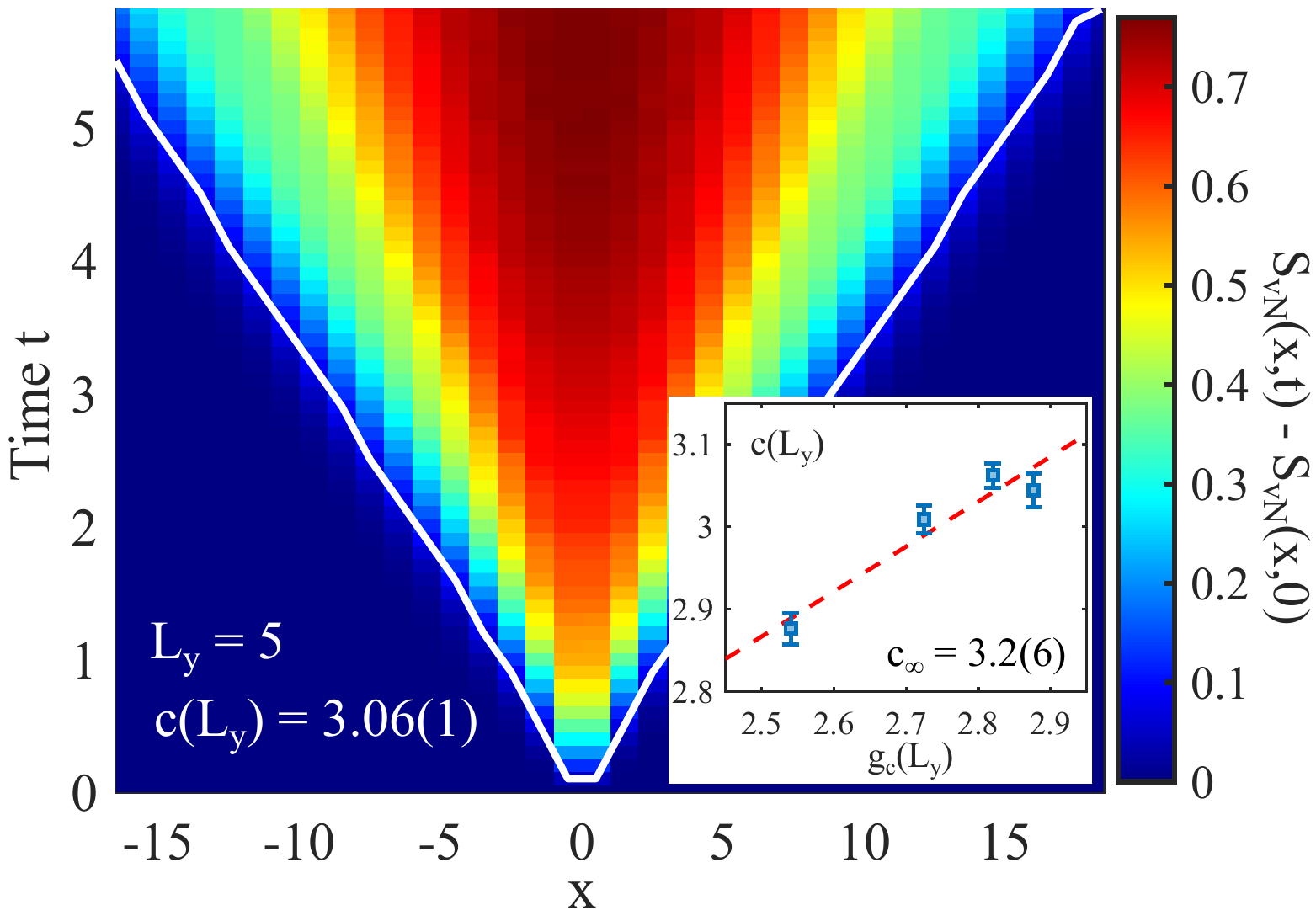}
    \caption{The ground state of the 2D-TFI for $L_y=5$ at criticality ($g_c=2.8202$) is perturbed at site $(x,y)=(L_x/2+1,1)$ and is time evolved. The von Neumann entanglement entropy is calculated at every time step at the MPS bonds that bisect the cylinder along its length. A light cone appears, which allows the estimate of the speed of light by doing a linear fit to the white line, giving $c \approx 3.06(1)$. \textbf{Inset} The speed of light is shown for system sizes up to $L_y=6$ with error bars corresponding to the small range of velocities that reasonably represented the boundaries of the light cone. The red line is a linear fit to $c(L_y) = ag_c(L_y)+c_\infty$.}
    \label{fig:vCrit}
\end{figure}

The 2D-TFI model has a second order phase transition separating a ferromagnetic phase for $g<g_c$ and paramagnetic phase for $g>g_c$. The critical point of the Hamiltonian in Eq.~(\ref{eq:Hamiltonian}) is verified by performing a scaling collapse of the energy gap calculated with DMRG as a function of the transverse field, for system sizes up to $L_y = 6$. We find that our data agrees well with the known calculated critical field $g_c=3.04438$ and critical exponent $\nu=0.629971$ obtained through exact diagonalisation, quantum Monte Carlo, DMRG and experimental investigations~\cite{Henkel1988,duCroo1998,Hamer2000,Blote2002,ebadi2021,Pelissetto2002,Huang2020}, as shown in Fig.~\ref{fig:energyGap}.

Dynamical phase transitions have been studied in the ferromagnetic 2D-TFI model for homogeneous quenches of the transverse field across the critical point. These studies reveal that local spin excitations are the energetically dominant quasiparticles~\cite{Hashizume2022} and exhibit Kibble-Zurek scaling of the correlations for fast quenches across the critical point~\cite{zurek1985,zurek1993,zurek1996,Schmitt2022}. The latter also revealed that the adiabatic limit is reached for slow quenches, such that the ground state of the 2D-TFI can be prepared in a time scaling quadratically with the linear size of the system $L$. The critical point has dynamical critical exponent $z=1$, such that emergent low-energy theory has relativistic linearly dispersing modes. In this work, the velocity of excitations is estimated from the time evolution of a perturbation on the lattice. The bipartite von Neumann entanglement entropy shows the emergence of a light cone from which a velocity can be extracted, as seen in Fig.~\ref{fig:vCrit}. Indeed, the white line in Fig.~\ref{fig:vCrit} represents a light cone with speed of light $c\approx 3.06(1)$. The system size scaling of the speed of light is observed to be approximately linear in $g_c(L_y)$ as can be seen in the inset of Fig.~\ref{fig:vCrit}. A linear fit gives a speed of light in the thermodynamic limit of $c_\infty \approx 3.2(6)$  

\section{Doppler Cooling in the 2D-TFI model}
\label{sec:dopplerCooling}
\begin{figure}
    \centering
    \includegraphics[width=\linewidth]{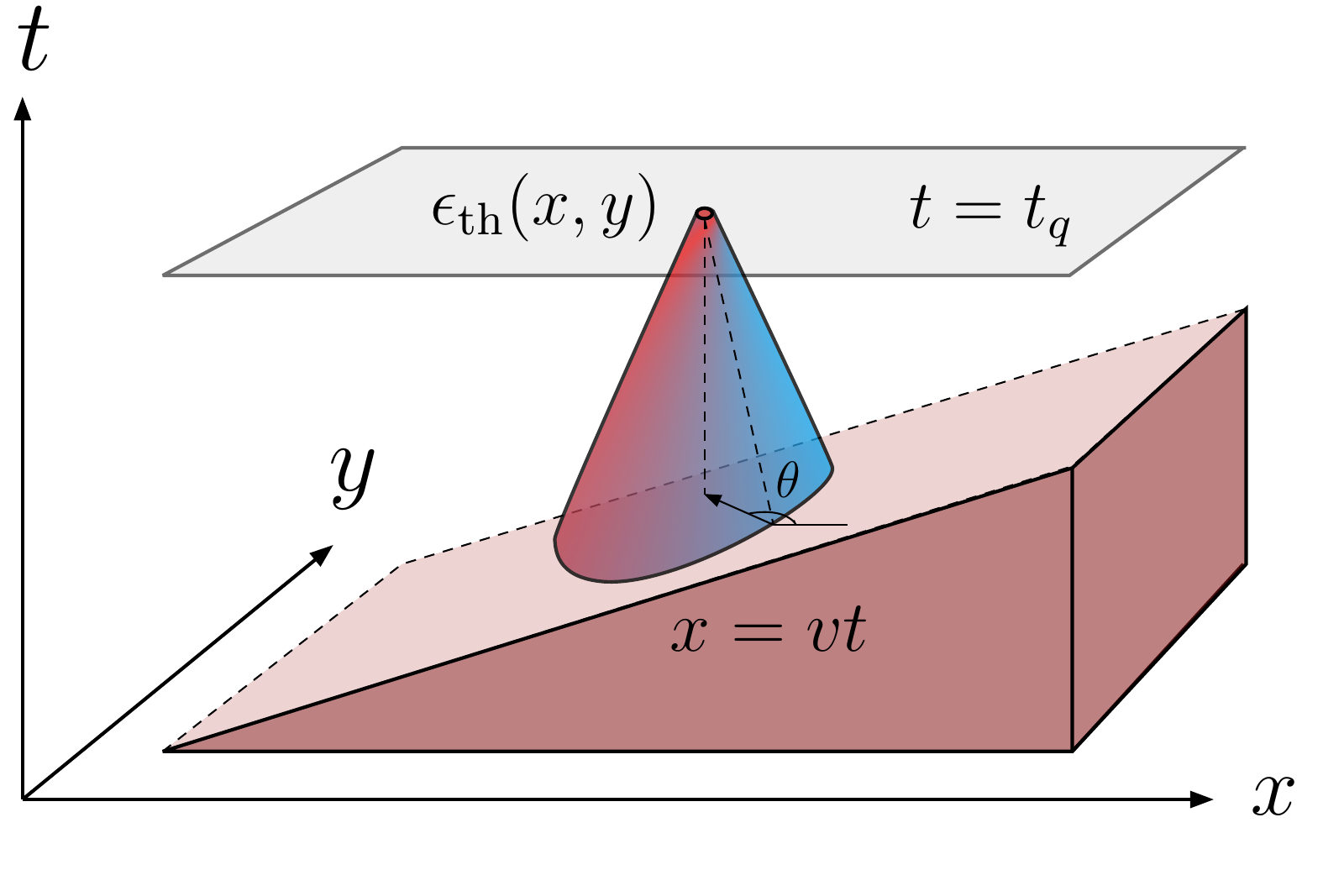}
    \caption{The spatial distribution of energy can be theoretically estimated by integrating over excitations emanated from the quench front whose populations are Doppler shifted according to the angle of emission $\theta$; see Eq.~(\ref{eq:energyTheory}). Here blue indicates colder waves excited against the quench front, and red indicates hotter waves copropagating with the quench front.}
    \label{fig:coolingexpfig}
\end{figure}

\begin{figure}
    \centering
    \includegraphics[width=\linewidth]{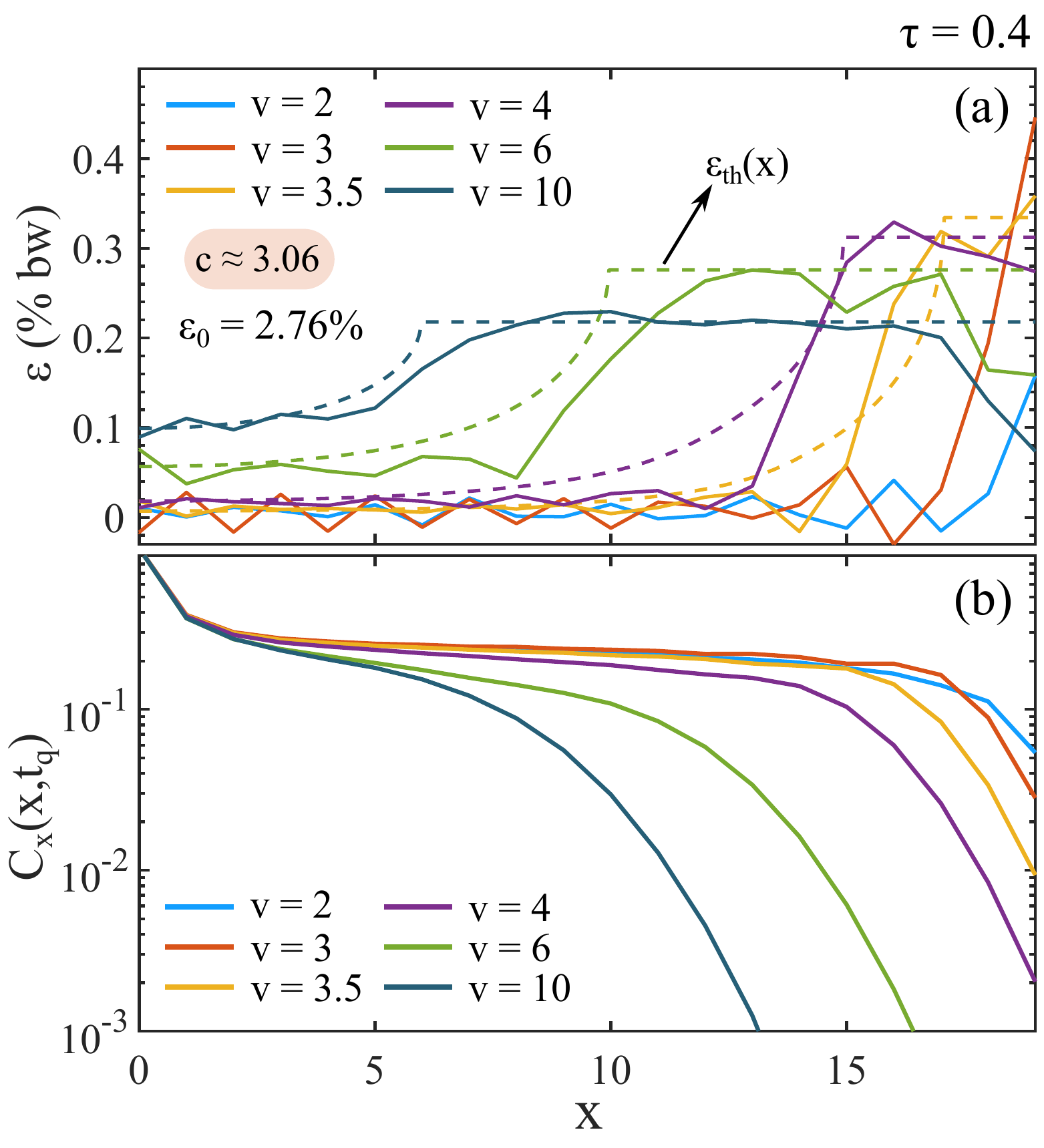}
    \caption{(a) Spatial distribution of the energy density during a spatiotemporal quench at time $t_q=\frac{1}{2}(L_x-1)/v$ shows a heatwave picture consistent with two-dimensional calculations for quenches in free bosonic theory. The observed  oscillatory behaviour potentially corresponds to UV modes being excited non-chirally. The dashed lines correspond to the theoretical curves for the energy density at time $t_q$ after a spatiotemporal quench in 2D free fermions. (b) The spin correlations at time $t_q$ show correlation lengths that are consistent with the ``cold" and ``hot" regions that explain the energy density in the heatwave picture.}
    \label{fig:heatwave}
\end{figure}

The heatwave picture developed in Ref.~\cite{agarwal2017} summarizes the relativistic cooling mechanism for superluminal quench fronts presented in this section. Modes are excited corresponding to a Doppler shifted temperature that depends on the angle $\theta$ at which they are emitted relative to the front trajectory. In one dimension, this creates a well demarcated hot and cold regions depending on whether that region is illuminated by waves traveling against the quench front (cold) or with the quench front (hot). In two and higher dimensions, waves can be excited in more directions and are Doppler shifted by a factor $\eta (\theta)$ that is dependent on the angle with respect to the quench front at which they are radiated. The cooling effect is expected to persist in the higher dimensional case as well since even waves that are excited orthogonal to the quench front are Doppler shifted downwards, by a Doppler factor $\eta (\pi/2) = \gamma$, where $\gamma > 1$ is a Lorentz dilation factor. In general, we expect a spatially varying energy density profile, as indeed observed in our simulations; see Fig.~\ref{fig:heatwave}(a).  In particular, we expect a ``hot" region for $ct<x<vt$ populated by blue-shifted modes and a ``cold" region for $x \ll ct$, with the energy density continuously varying between the two extremes. The ``hot" and ``cold" regions are also identified in the spin correlations. The colder regions correspond to slow, power-law decay of correlations, while the hot regions exhibit exponentially decaying correlations with a small correlation length, as shown in Fig.~\ref{fig:heatwave}(b).

For free relativistic fermions, as pertains to the nearest-neighbour one dimensional critical TFI model, one can perform detailed calculation of these mode populations~\cite{bernier2022spatiotemporal}. In the two dimensional case, the 2D TFI model cannot be mapped to free fermions; the natural interpretation is instead in terms of interacting bosons. For free bosons, the result for the mode population in two dimensions and for $\tau = 0$ is given by~\cite{agarwal2018}
\begin{equation}
    N_\theta (\textbf{k}) \approx
        \begin{cases}
        \frac{m}{4 \eta (\theta) \omega_k} & \; \; \mbox{for} \; \; \eta (\theta) \omega_k \ll m \\
        e^{-2 \eta(\theta) \omega_k m} & \; \; \mbox{for} \; \; \eta(\theta) \omega_k \gg m
        \end{cases} 
\end{equation}
where $\omega_k=c|\textbf{k}|$, $m$ is the mass associated with the initial gap, 
$\gamma = 1/\sqrt{1-\beta^2}$ is the Lorentz factor with $\beta=c/v$ and $\eta(\theta) = \gamma(1-\beta\cos\theta)$ is the relativistic Doppler factor associated with modes propagating at an angle $\theta$ relative to the quench front trajectory. Here, $\eta(\pi) \equiv \eta = \sqrt{\frac{1+\beta}{
1-\beta}}$ is associated with counterpropagating modes, $\eta(0)=1/\eta$ is associated with copropagating modes and transverse modes receive a factor $\eta(\pi/2)=\gamma$. 
As $v\rightarrow c^+$, the Doppler factor diverges and the region left in the wake of the quench front is left completely unexcited.

The energy density associated with modes emitted at angle $\theta$ during a quench is obtained by integrating the population of bosons carrying energy $\omega_k$ at angle $\theta$ over the momentum $k$. This yields
\begin{equation}\label{eq:energyTheory}
\epsilon(\theta) \propto \int_0^{m/\eta(\theta)} \omega_\mathbf{k}N_\mathbf{k} k dk \propto \frac{m}{L^2}\frac{1}{\eta(\theta)^3},
 \end{equation}
where $L$ is the linear length of the system.

For spatiotemporal quenches in the 2DTFI, the analysis is complicated on multiple fronts---i) the low energy theory should correspond to interacting bosons and ii) there is a natural UV cutoff associated with the finiteness of the local Hilbert space. The latter particularly implies that the UV modes deviate from the perfect linear dispersion and will generally not get excited in a chiral fashion. Therefore, their population must be controlled. We enable this by using a finite quench time $\tau$. Assuming the UV cutoff $\Lambda > 1/\tau$, this effect can be introduced by approximately changing the cutoff of the integral in Eq.~(\ref{eq:energyTheory}) to $\text{min} \left[ m/ \eta(\theta), 1/\tau \right]$. This is particularly relevant for waves that are Doppler shifted to higher energies, for which the energy density will scale as $\sim 1/\eta(\theta)$. 

\begin{figure}
    \centering
    \includegraphics[width=\linewidth]{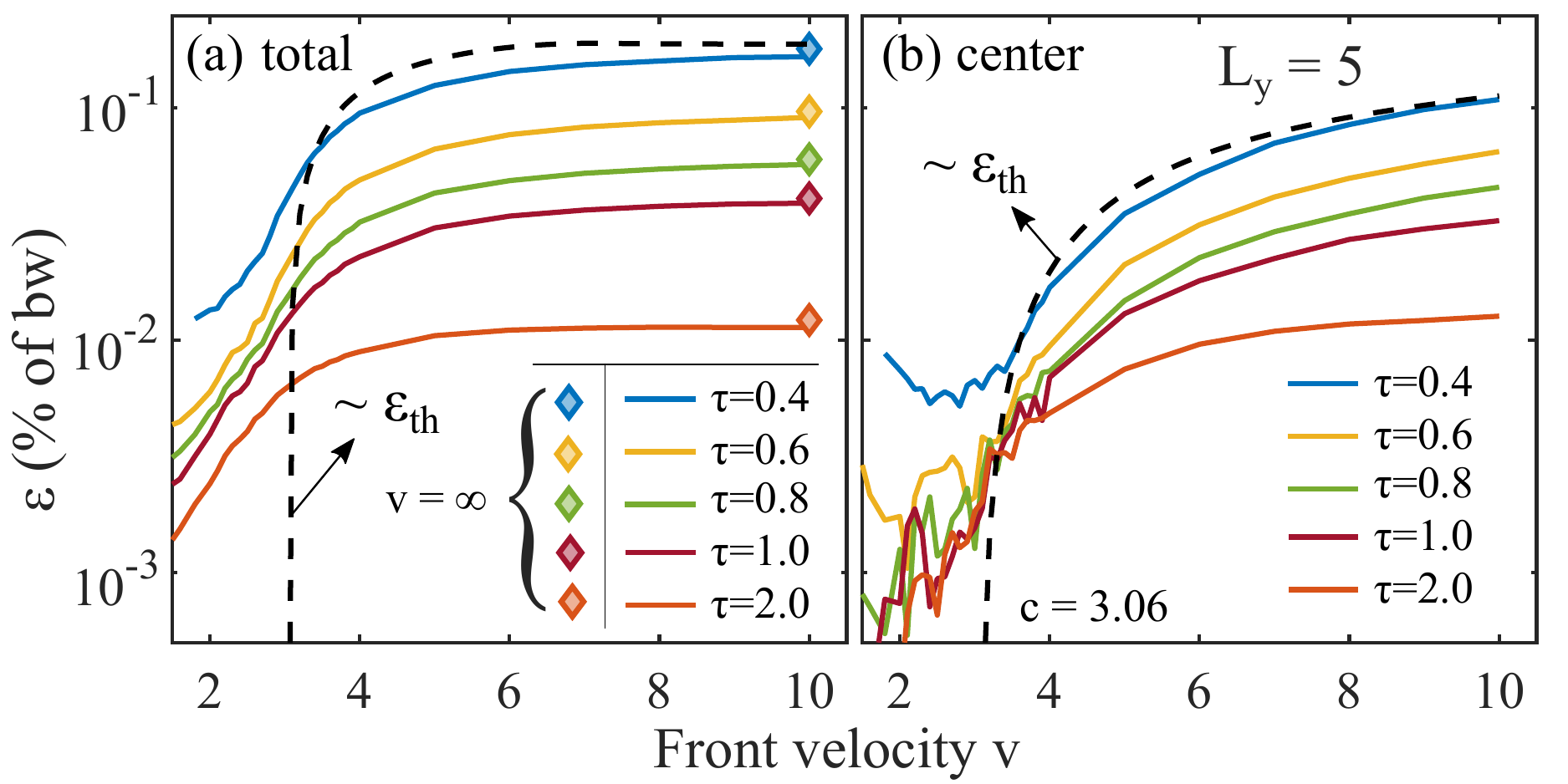}
    \caption{(a) The total energy density in a system of size $L_y=5$ at time $t_q$ is shown for different $\tau$. The energy density $\int \epsilon(\theta)d\theta$ averaged over the entire system is plotted as a black-dotted line for $c=3.0625$. For every $\tau$, the energy density at the end of a uniform quench ($v=\infty$) is plotted using a diamond marker. (b) The average energy density in the center of the cylinder in a region of size $L_y\times 2L_y$ is shown. The expected energy density $\int \epsilon(\theta)d\theta$ averaged over the same region is plotted as a black-dotted line.}
    \label{fig:energyV}
\end{figure}

One can now compute the energy density at any point in space by integrating over the contributions from waves emitted at all different angles from the quench front and arriving at that point in space at some fixed time; see Fig.~\ref{fig:coolingexpfig} for a visual description of this computation. In all our calculations, we compute the spatial energy distribution at time $t = t_q = L_x/2v$ at which point the quench front reaches the ends of the system. We denote the corresponding result for the energy density $\epsilon_{\text{th}} (x)$; we plot these results in Fig.~\ref{fig:heatwave}(a). In addition, we compute the average energy density in the total system at the end of the quench, and the average energy density in a finite region of length $2L_y$ in the $x$-direction and $L_y$ in the $y-$direction centered at $x = 0$. We show these results in Fig.~\ref{fig:energyV}. 

Comparing with numerical results, we see reasonably good agreement of the data for the spatial distribution of the energy density shown in Fig.~\ref{fig:heatwave} (a) in the region $x < ct$ where we anticipate cooling and thus less of an issue with UV modes. The theoretical curves are normalized to match the energy density for $v=10$ at time $t_q$ in the center of the hot region. In the hot regions $ct<x<vt$, the energy density is lower than the heatwave prediction due to the finite time scale $\tau$ preventing the excitation of blue-shifted high momentum modes, along with fact that the calculations for free bosons assume infinite dimensional local Hilbert space (which theoretically allows for an unboundedly large energy density), which is not applicable to the spin model we consider. The presence of hot and cold regions is also seen in the decay of the spin-spin autocorrelations; see Fig.~\ref{fig:heatwave} (b), with correlations showing slower decay in the colder regions and fast exponential decay as the hot region is approached. 

In Fig.~\ref{fig:energyV} (a) and (b), we show the total and central region's average energy density as a function of the quench front velocity $v$, for various quench parameters $\tau$. The theoretical curve appearing from free boson calculation for $\tau = 0$ is plotted as a dashed line for reference. The numerical results generally agree with the theoretical curve for small $\tau$, where adiabatic effects owing to this parameter may be ignored, and larger front velocities $v \gg c$. As $v \rightarrow c^+$, we note that the numerical results deviate from the theoretical predictions which predict stronger cooling. This can be explained by the fact that at velocities closer to the speed of light, UV modes are more strongly excited; since these modes deviate from linear dispersion which belies relativistic Doppler cooling physics, the cooling becomes less efficient. The results for the velocities closer to the speed of light appear to agree better for larger $\tau$, where UV modes should be more suppressed. We note that at small $\tau$, we also observe a minimum in the central region's energy density as a function of quench velocity---we expect cooling to become inefficient in the subluminal case for similar reasons to the superluminal case~\cite{bernier2022spatiotemporal}.

\begin{figure}
    \includegraphics[width=\linewidth]{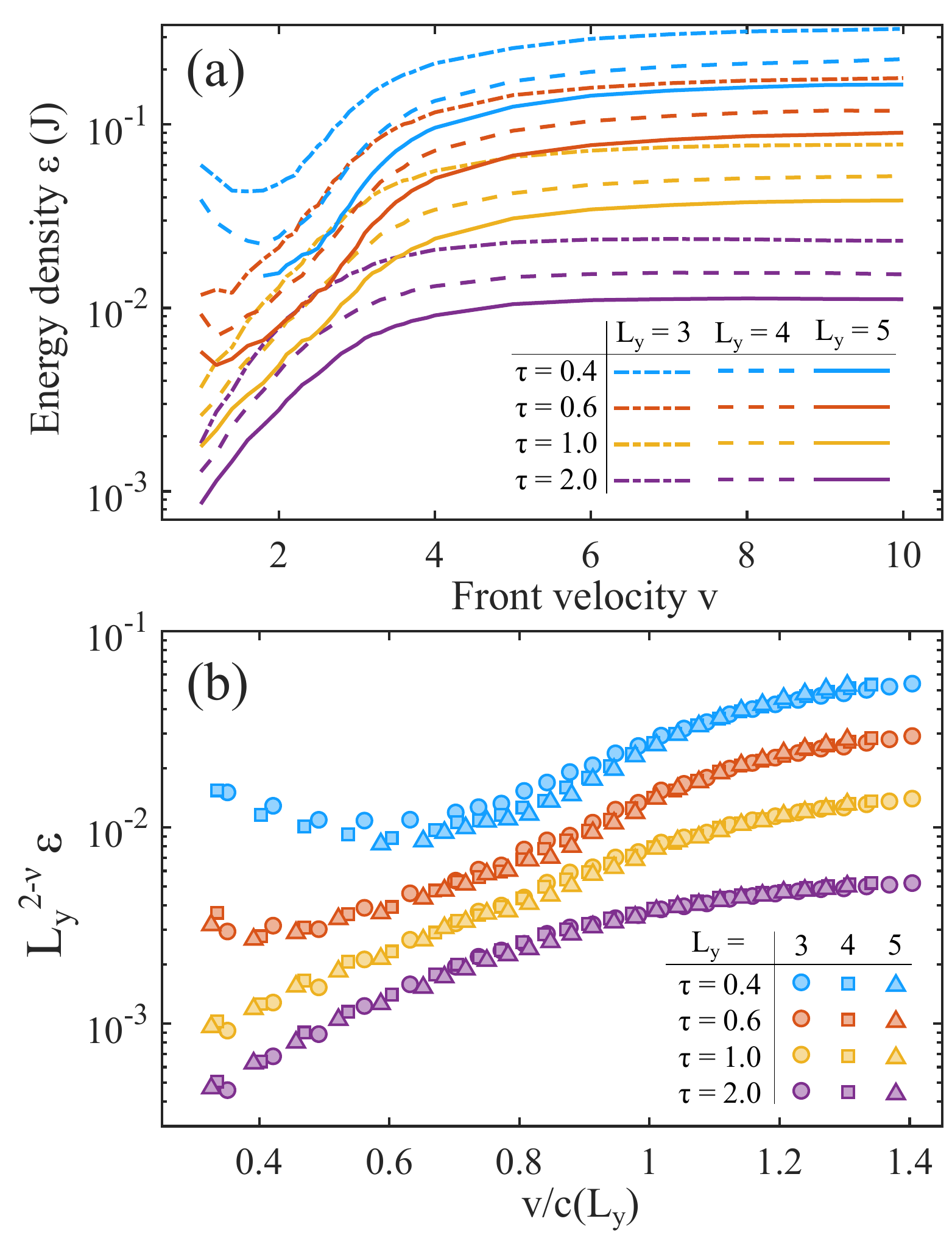}
    \caption{(a) The energy density at time $t_q$ is shown for different $L_y$ and $\tau$. (b) A scaling collapse of the energy density according to the critical scaling $L_y^{-(2-\nu)}$ found in Appendix~\ref{sec:crit2DTFI}.} 
    \label{fig:energyScaling}
\end{figure}

\begin{figure}
    \includegraphics[width=\linewidth]{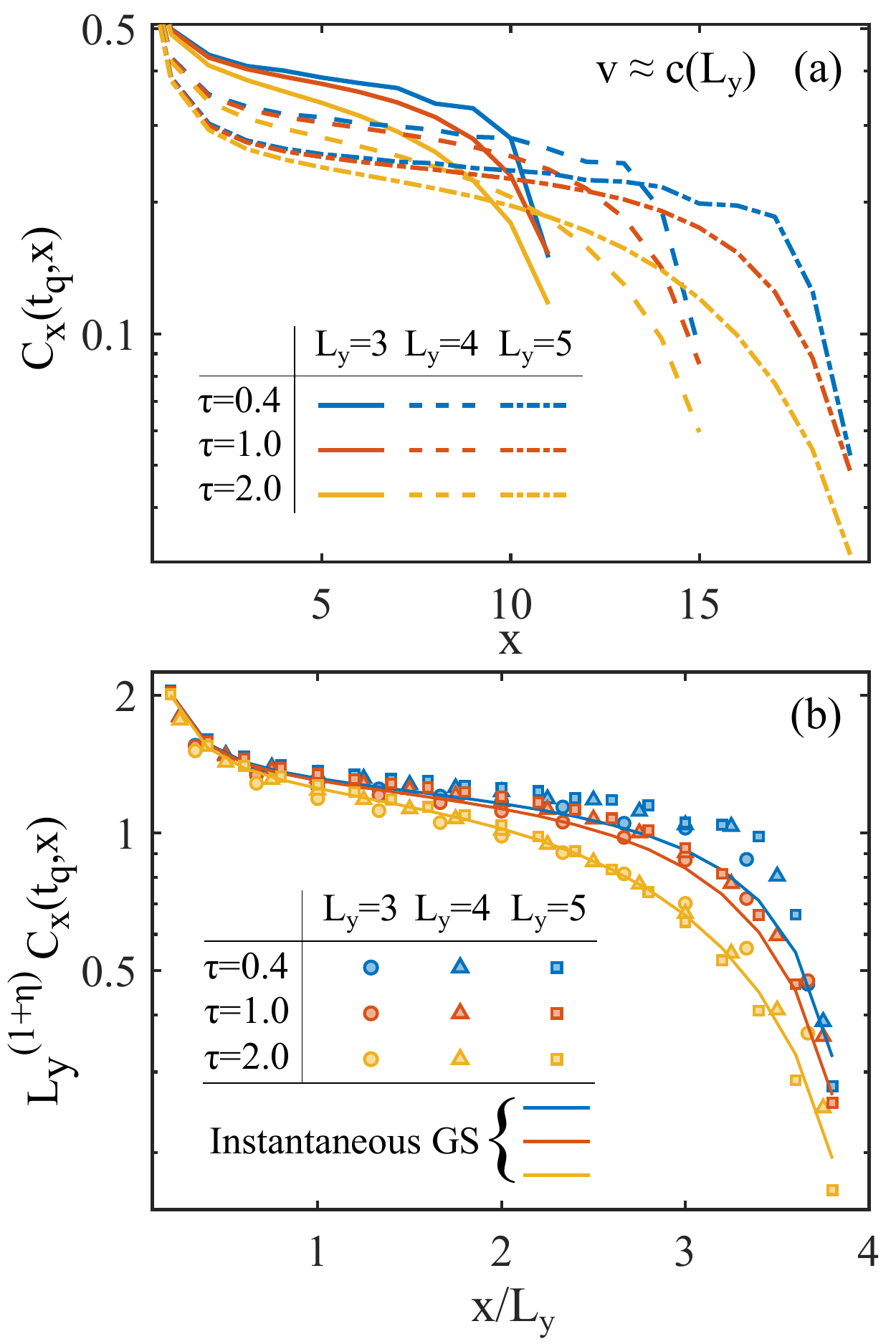}
    \caption{(a) The spin correlations for $v\approx c(L_y)$ are shown at time $t_q$ for various $\tau$ and system sizes $L_y$. (b) A scaling collapse of the spin correlations with $L_y$ is shown for different $\tau$. The critical exponent of the autocorrelator $2\Delta=1+\eta$, where we use the known critical exponent $\eta=0.629971$. The scaled autocorrelator of the instantaneous ground states (GS) at $t_q$ are plotted for $L_y = 5$.}
    \label{fig:spinCorr}
\end{figure}

\begin{figure}
    \includegraphics[width=\linewidth]{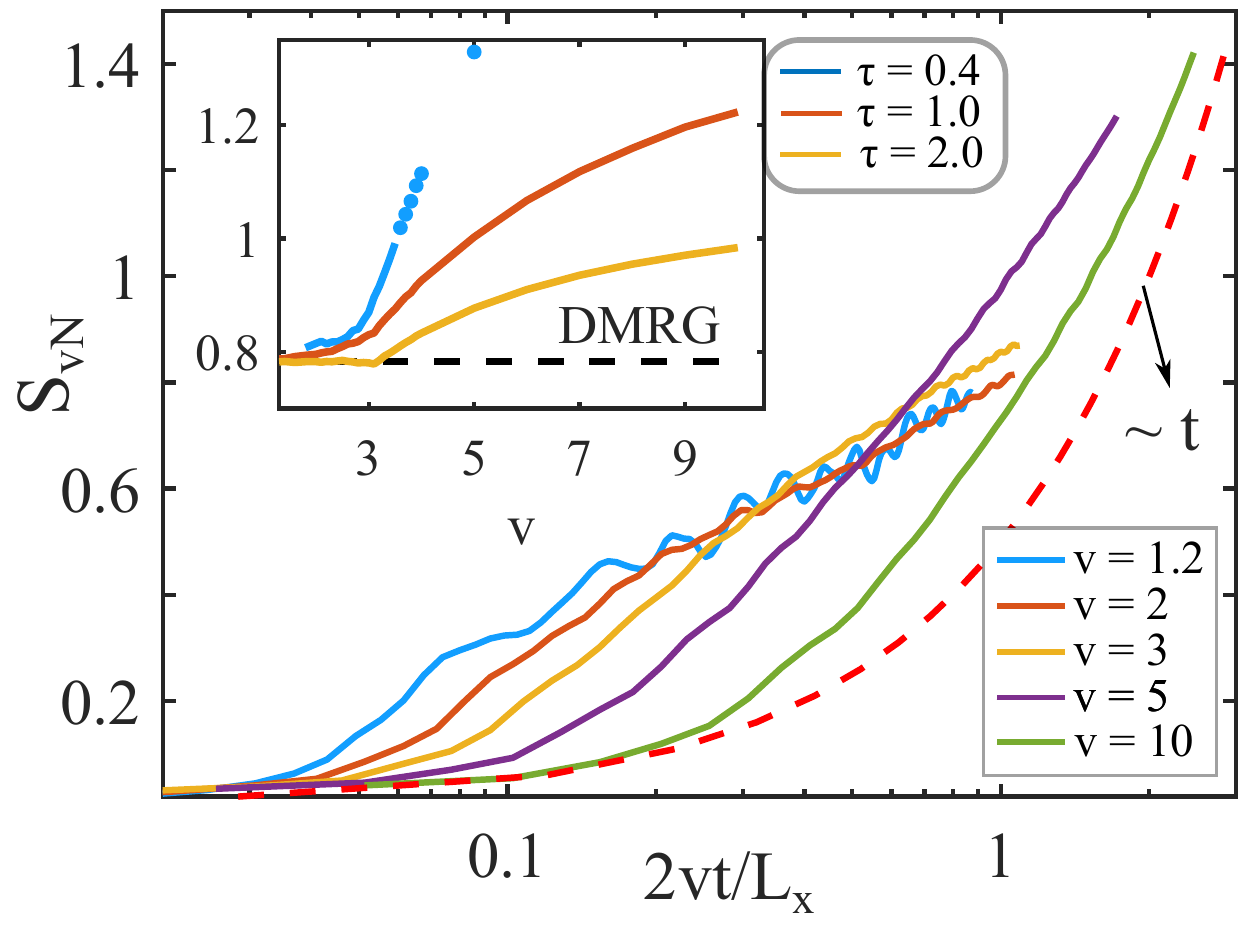}
    \caption{The growth of the von Neumann entanglement entropy during the quench is shown for $\tau=0.4$ and different quench velocities. The entanglement entropy shows approximately logarithmic growth for velocities $v\leq3$. The quench ends at $2vt/L_x=1$, but the data shows the entanglement entropy keeps increasing until $v=L_x/2c$, after which it saturates and oscillates (data not shown). Inset: the entanglement entropy at time $t = L_x/2c$ is shown for different $\tau$ as a function of front velocity when computationally accessible. The circles represent data at $\chi = 256$. The black-dotted line shows the entanglement calculated at criticality using DMRG. }
    \label{fig:svn} 
\end{figure}

Further insight into the critical properties of the final state is obtained by performing a scaling collapse with system size scale $L_y$ of the energy density and spin-spin correlations. As seen in Fig.~\ref{fig:energyScaling} (b), near the speed $c(L_y)$, the total energy density appears to scale as $L_y^{-(2-\nu)}$ where the critical exponent agrees with that seen at DMRG simulations of the system near the critical field. The $L_y$ scaling of the spin correlations at $v=3$ is tested for different $\tau$ as shown in Fig.~\ref{fig:spinCorr}(a). Quantitatively, we examine the autocorrelator using the following ansatz
\begin{equation}\label{eq:equiSC}
    C_x(x,t_q) = \langle \sigma_0^x \sigma_{r}^x \rangle = L_y^{-2\Delta} F_C\left(\frac{r}{L_y}\right),
\end{equation}
which identifies the scaling of the correlation length with the system size $L_y$ and critical exponent $2\Delta = 1+\eta$, where $\eta=0.0363$ is the known critical anomalous dimension of the 2D-TFI \cite{Blote2002,Schmitt2022}. We note that the autocorrelation function shows good collapse with $L_y$ in the colder region where critical scaling is expected to be seen, and for larger $\tau$, as seen in Fig.~\ref{fig:spinCorr}(b). The scaled autocorrelator of the ground state for $L_y=5$ obtained by DMRG (labeled as the instantaneous GS) is plotted as a solid line against that obtained from the quench in Fig.~\ref{fig:spinCorr}(b). It reveals that as $\tau$ is increased, the system is quenched more efficiently to the ground state of the critical theory. This scaling collapse shows that near $v=3$, our quench protocol indeed produces correlations close to the critical correlations of the 2D-TFI model.




Finally, we also study the growth of the bipartite von Neumann entanglement entropy, with $x = 0$ being the position of the partition.  For the quench, the entanglement entropy ($S_\text{vN}$) appears to increase much more slowly in time for $v \approx c$ as compared to the expected linear growth for a homogeneous quench~\cite{calabrese2005,calabrese2009}; see Fig.~\ref{fig:svn}. For $v \gg c$, we indeed observe a much stronger growth in entanglement entropy that is almost linear in $t$. We can interpret the result as follows. Entanglement between the two halves of the system increases as one of the excitations emitted in the direction of the partition $x = 0$ from the quench front crosses the partition. Assuming there's a maximum speed $c$ setting causal speed limit on information flow in this system, the region $x \ge x_c = c t / ( 1 + c/v)$ will generically be causally decoupled from the region $x < 0$ and will not contribute to entanglement in the system. Assuming next that the region $x < x_c$ is found at the zero temperature critical state of the 2DTFI model, we may expect the entanglement entropy, $S_{\text{vN}}$, to scale as $L_y \text{log} x_c \propto \log t$. If the region $x < x_c$ is heated to a finite temperature, then we can expect volume law entanglement $S_{\text{vN}} \sim L_y x_c \propto t$. Our numerical findings thus suggest that for $v \lesssim c$, one approaches very close to the ground state of the system. Our results in fact appear to go a step further and suggest (see inset) that the entanglement entropy at the point where excitations produced at $t=0$ reach the system's boundary, saturates to the ground state entanglement entropy (as obtained by DMRG) in the 2DTFI model for the \emph{full} system. We do not fully understand this result.

These results demonstrate that spatiotemporal quenches can efficiently produce critical ground states in two-dimensional systems. In light of previous studies \cite{sinha2020,bernier2022spatiotemporal}, we believe this quench protocol will be efficient in current quantum simulators with short-range interactions, making it immediately relevant to the field of condensed matter physics. In particular, ultracold Rydberg atom are a setup where such a protocol could be used to realize ground states of interacting Hamiltonians, starting from ground states that are product states (as is the case with large transverse fields). The ability to individually address the fields on each atom in this setup should allow for the realization of spatiotemporal quenches such as the ones we propose. Trapped ion arrays are another promising avenue for the realization of effective spin models but generally exhibit long-range hopping and interactions decaying as $1/r^\alpha$. In two-dimensions, the tightest Lieb-Robinson bounds predict that linear light-cones $t \sim r$ (indicating relativistic low-energy dynamics) exist in these systems only for $\alpha > 5$~\cite{kuwahara2020}, which reduces the range of application of this particular protocol to van der Waals interactions ($\alpha=6$) implementable in ultracold Ryberg atoms. It remains a matter of future investigation to determine if other quench protocols $v(t)$ can be implemented to efficiently prepare the critical ground states of 2D systems with long-range interactions.

\section{Conclusion}
\label{sec:conclusion}

This work investigates the efficiency of preparing critical ground states of the two-dimensional transverse field Ising (2D-TFI) model using spatiotemporal quenches. The study demonstrates that the protocol effectively leads to the rapid production of ground states, even in critical cases characterized by linearly dispersing modes and vanishing energy gaps. The results show that large sections of the system are left unexcited when the quench front velocity $v \approx c$, the emergent speed of light, resulting in critical ground state correlations. For smooth quench fronts that leave UV modes unexcited, critical scaling relations accompanied by a collapse of spin correlations show that the correlation length saturates and decays on the length scale $L_y$ when $v\rightarrow c$. For general $v>c$, a heatwave picture emerges where one obtains hot and cold regions in the system populated by excitations emanating from the quench front. The optimal quench protocol also shows a nearly logarithmic growth of the von Neumann entanglement entropy.

The ability to prepare ground states in strongly correlated systems is crucial for exploring the properties and behavior of quantum materials. We have not here addressed the question of boundaries which inevitably will play a role in the final cooling outcome. We assume that since much of the hot region is unentangled with the other half of the system, it can be effectively decoupled without much gain in energy but this requires more careful investigation. Exploring the applicability of spatiotemporal quenches in models with longer range interactions, as relevant in trapped ion simulators, and the study of spatiotemporal quenches with an accelerating/decelerating quench front also represent interesting possibilities for further inquiry. Finally, many of these protocols will also have to be applied in concurrence with a number of dynamical decoupling protocols~\cite{choi_robustdynamicH,agarwaldynenh2020,martinprx2020,KasperDDGauge2023} needed to realize effective Hamiltonians of interest in these artificial atom/ion setups. 

\begin{acknowledgments}
The authors acknowledge useful discussions with several previous collaborators on related work. SB acknowledges support from an FRQNT graduate scholarship. KA acknowledges support of the Material Sciences and Engineering Division, Basic Energy Sciencs, Office of Science, US-DOE, and previous support from the NSERC Discovery Grant and an INTRIQ team grant from the FRQNT.
\end{acknowledgments}

\appendix

\section{Critical Properties of the 2D-TFI Model on a Cylinder}\label{sec:crit2DTFI}

\begin{figure}
    \includegraphics[width=\linewidth]{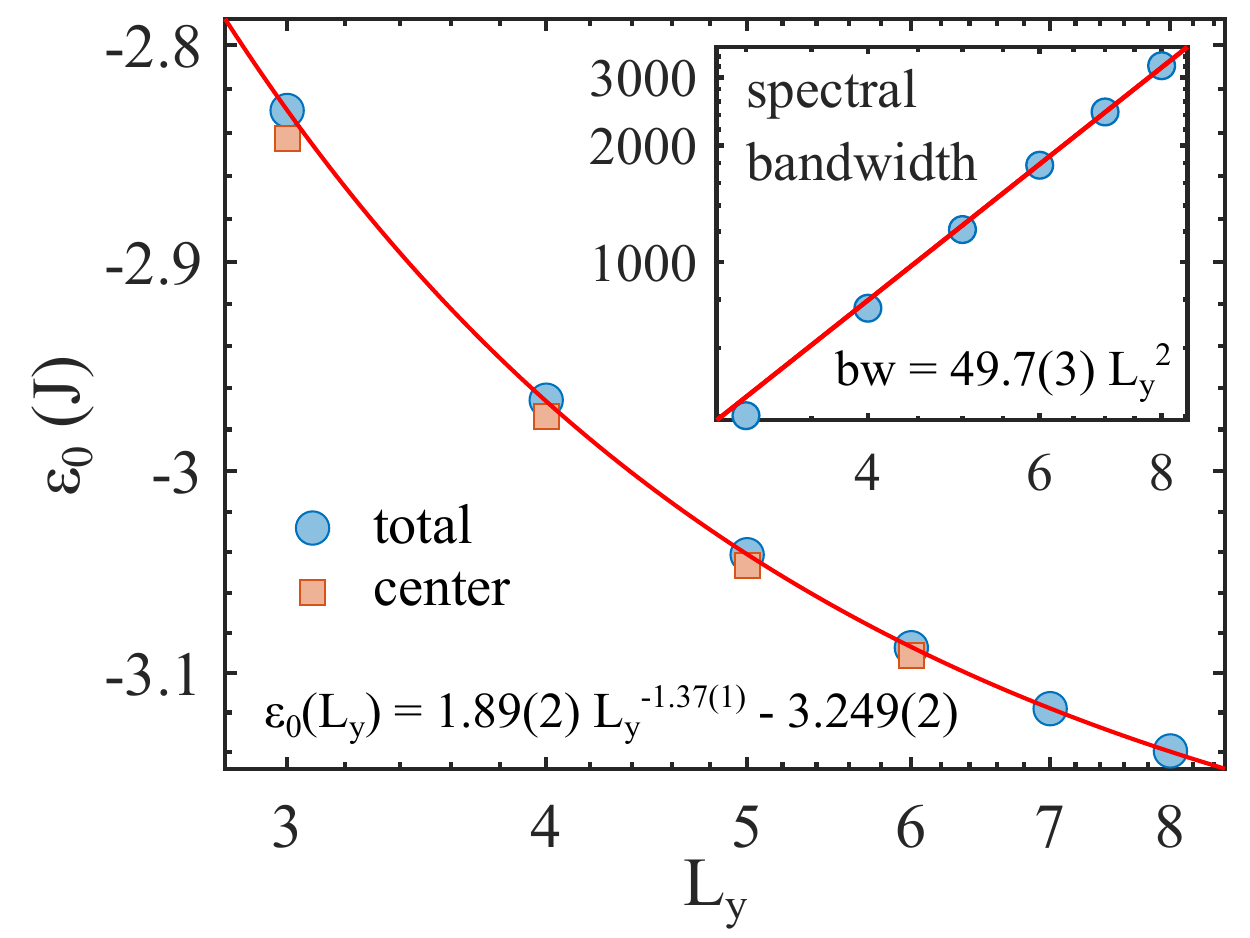}
    \caption{The ground state energy density of the 2D-TFI model is shown at criticality over the entire system and within the central region of the system. A fit is done to the data showing the existence of a subleading term scaling as $L_y^{-1.37}$.}
   \label{fig:critEnTotal} 
\end{figure}

We study the critical properties of the 2D-TFI model with y-periodic boundary conditions to confirm the scaling exponents of the energy density and auto-correlation function. We begin by computing the ground state energy at criticality for different system sizes up to $L_y = 8$. The critical transverse field at each $L_y$ is determined by using the scaling collapse in Fig.~\ref{fig:energyGap}, giving
\begin{equation}
    g_c(L_y) = 3.04438 + a L_y^{1/\nu},     
\end{equation}
where we found numerically that $a = -2.88$. At these values of $g_c(L_y)$, the critical ground state energy density is shown in Fig.~\ref{fig:critEnTotal}. The ground state energy density shows a significant scaling with $L_y$ that vanishes in the thermodynamic limit, an effect that can be attributed to irrelevant fields. Indeed, a fit to the energy density shows that the energy density decreases with a power law $L_y^{-1.37(1)}$. We understand this scaling using the following ansatz for the critical ground state energy
\begin{equation}
    E_0 = Q L^2 + b L_y^{\nu},
\end{equation}
where $\nu$ is the critical exponent. The energy density for a system of size $N=8L_y^2$ then scales as 
\begin{equation}
   \epsilon_0 = \frac{E_0}{N} = Q' + b' L_y^{-(2-\nu)},
\end{equation}
with the exponent $2-\nu \approx 1.37$ agreeing with the fitted exponent. The energy density in a section of size $2L_y \times L_y$ in the center of the system shows a very similar behaviour for $L_y \leq 6$. We were unable to calculate the local energy density and spin correlations at criticality for systems larger than $L_y = 6$.

We also calculate the spectral bandwidth at criticality, which we use to normalize the energy in Fig.~\ref{fig:heatwave}(a) and Fig.~\ref{fig:energyV}. The spectral bandwidth is shown in the inset of Fig.~\ref{fig:critEnTotal} and scales approximately as $\sim L_y^2$ at larger system sizes.

\begin{figure}
    \includegraphics[width=\linewidth]{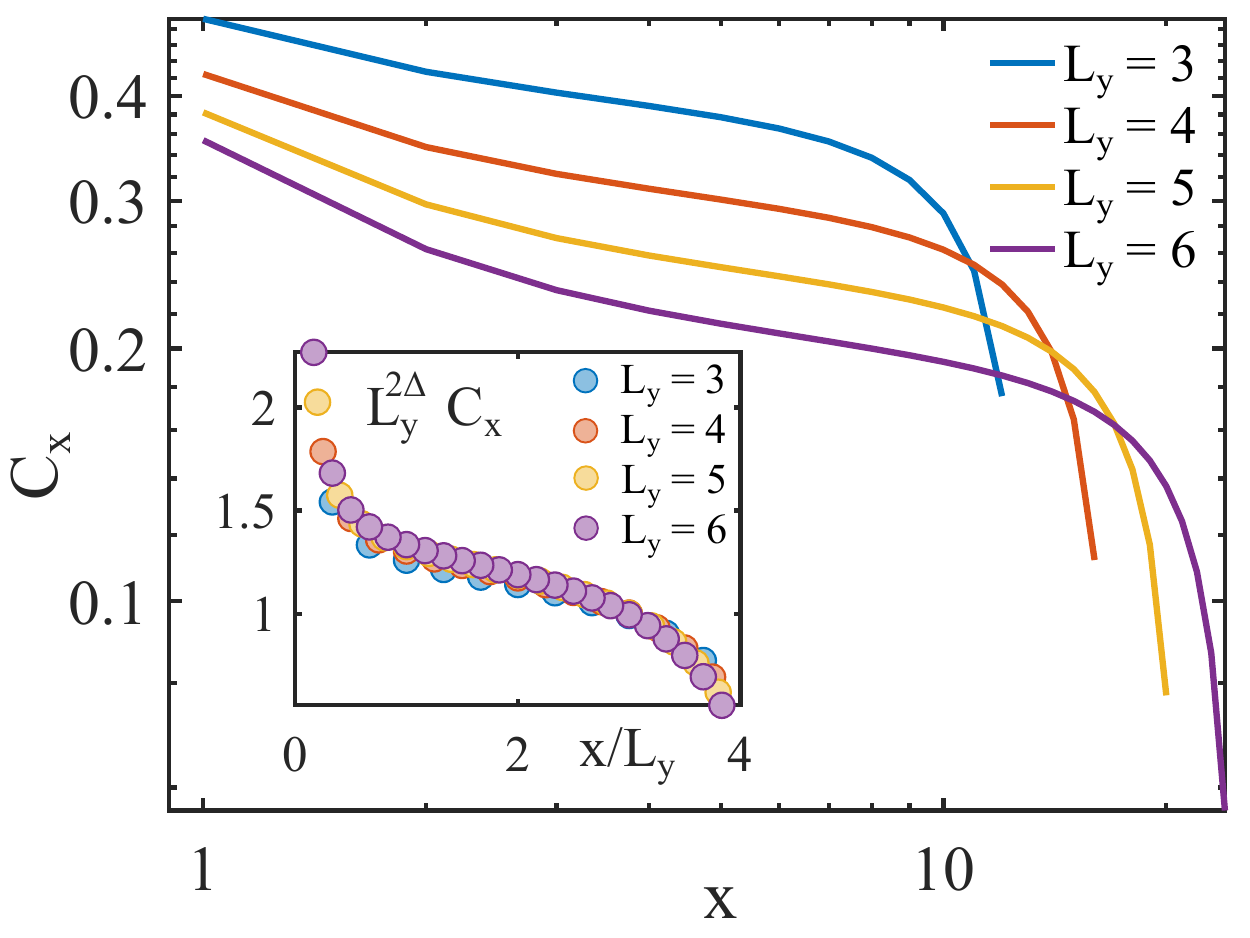}
    \caption{The critical ground state autocorrelator is shown for different system sizes. \textbf{\textit{Inset
    }} A collapse of the autocorrelators according to Eq.~\ref{eq:corrCritCollapse} is shown.}
    \label{fig:critCorr} 
\end{figure}

Next, we calculate the spin correlations along the x-axis of the system. The spin auto-correlations at criticality for system sizes up to $L_y = 6$ are shown in Fig.~\ref{fig:critCorr}. The correlation functions collapse well according to the critical scaling
\begin{equation}\label{eq:corrCritCollapse}
    C_x = L_y^{-2\Delta} F_c(x/L_y)
\end{equation}
where $2\Delta = 1 + \eta$, with $
\eta = 0.036298(2)$~\cite{Schmitt2022}. The scaling becomes increasingly better with larger $L_y$, due to a vanishing contribution to the expected critical behaviour in the thermodynamic limit, similar to the critical ground state energy.

We finally characterize the critical ground state entanglement entropy for different $L_y$. As in Ref.~\cite{Yu2008}, we find that the von Neumann entanglement entropy grows linearly with the area of the system $L_y$ with a subleading logarithmic correction. In Ref.~\cite{Yu2008}, the subleading term is obtained in the two-dimensional random TFI model as arising from an interaction of spin clusters with the boundaries of the system. We can expect that the presence of the logarithmic term in our system is also due to boundary interactions.

\begin{figure}
    \includegraphics[width=\linewidth]{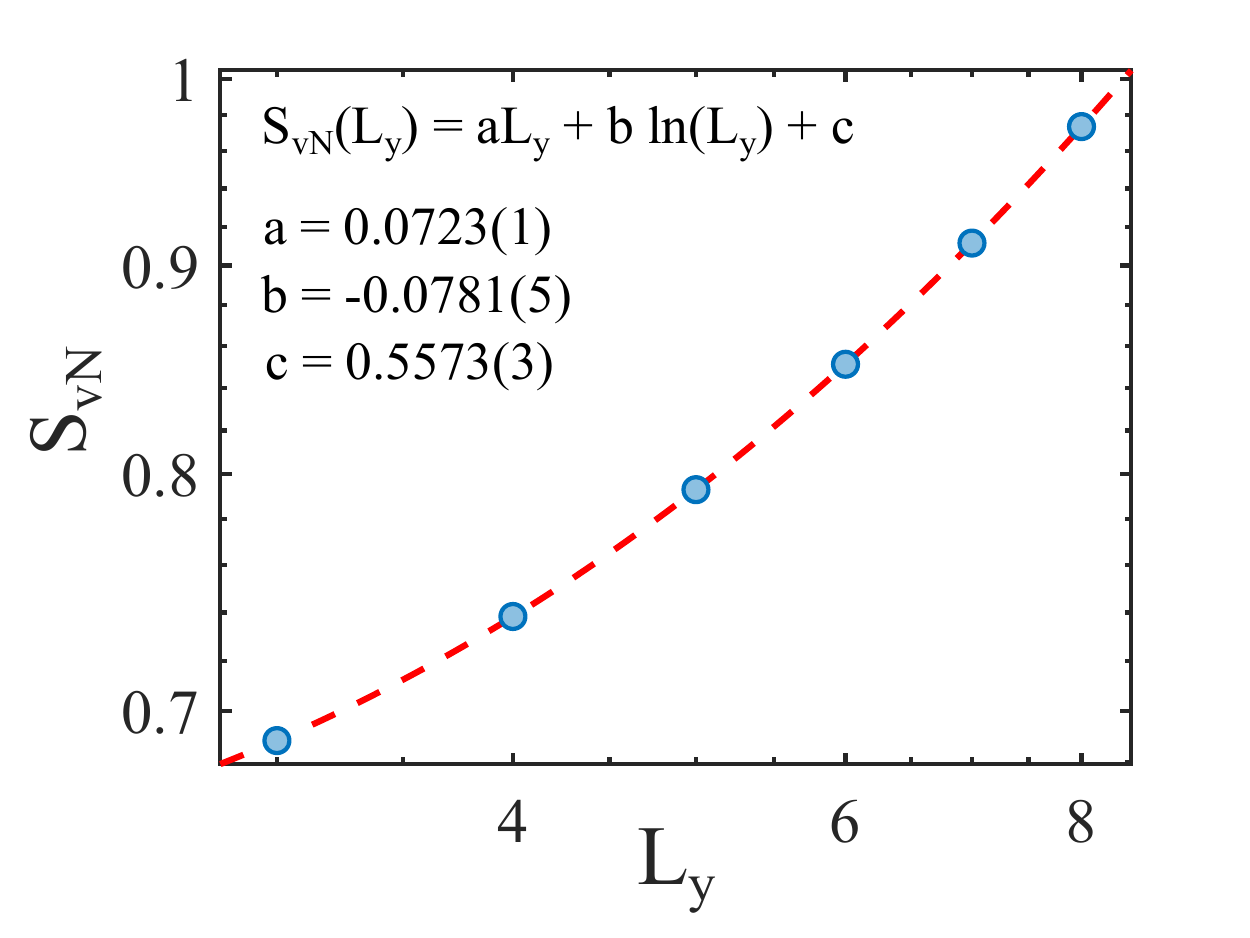}
    \caption{The von Neumann entanglement entropy computed in the center of the system is shown. A fit to $S_\text{vN} = a L_y + b \ln(L_y) + c$ is done, showing great agreement with the data and precise estimates of the fit parameters.}
    \label{fig:critSvn} 
\end{figure}

The data presented in this section motivate the use of $L_y = 5$ as a system of appropriate size such that the critical properties are close to those in the thermodynamic limit. In summary, at $L_y = 5$, the critical transverse field is given by $g_c \approx 2.8202$, with a ground state energy density  $\epsilon_0 = -3.041$ J,  deviating slightly from the fitted thermodynamic value of $-3.249(2)$ J. While the critical properties of systems with $L_y = 6$ were accessible, time evolution using fourth order TDVP did not converge due to the large size of the system and the larger bond dimension of the MPO representing the Hamiltonian. Even for smaller lengths $L_x = 4 L_y$, such that the system size and quench times are halved, the time evolution did not converge for every front velocity for system sizes above $L_y = 5$.

To obtain data for larger $L_y$ (comparable to Ref.~\cite{Schmitt2022}), we would need a parallel implementation of the algorithm or would need to decrease the aspect ratio of the system to $L_x = L_y$ or $L_x = 2L_y$. In that case, a heatwave picture similar to Fig.~\ref{fig:heatwave}(a) is difficult to obtain because the size of the region $x<ct = \frac{cL_x}{2v}$ would not be clearly distinguishable for a large range of parameters and system sizes. For example, for the largest quench velocity presented in this paper ($v=10$), the cold region $x<ct$ is distinguishable only above $L_y = 10$. Since the system is highly excited within the region $ct<x<vt$, the MPS bond dimension increases too quickly for such large systems to be simulated.

\section{Total energy density scaling at \texorpdfstring{$\tau=0.4$}{tau}} \label{sec:energyLyScalingLinear}

\begin{figure}
    \centering
    \includegraphics[width=\linewidth]{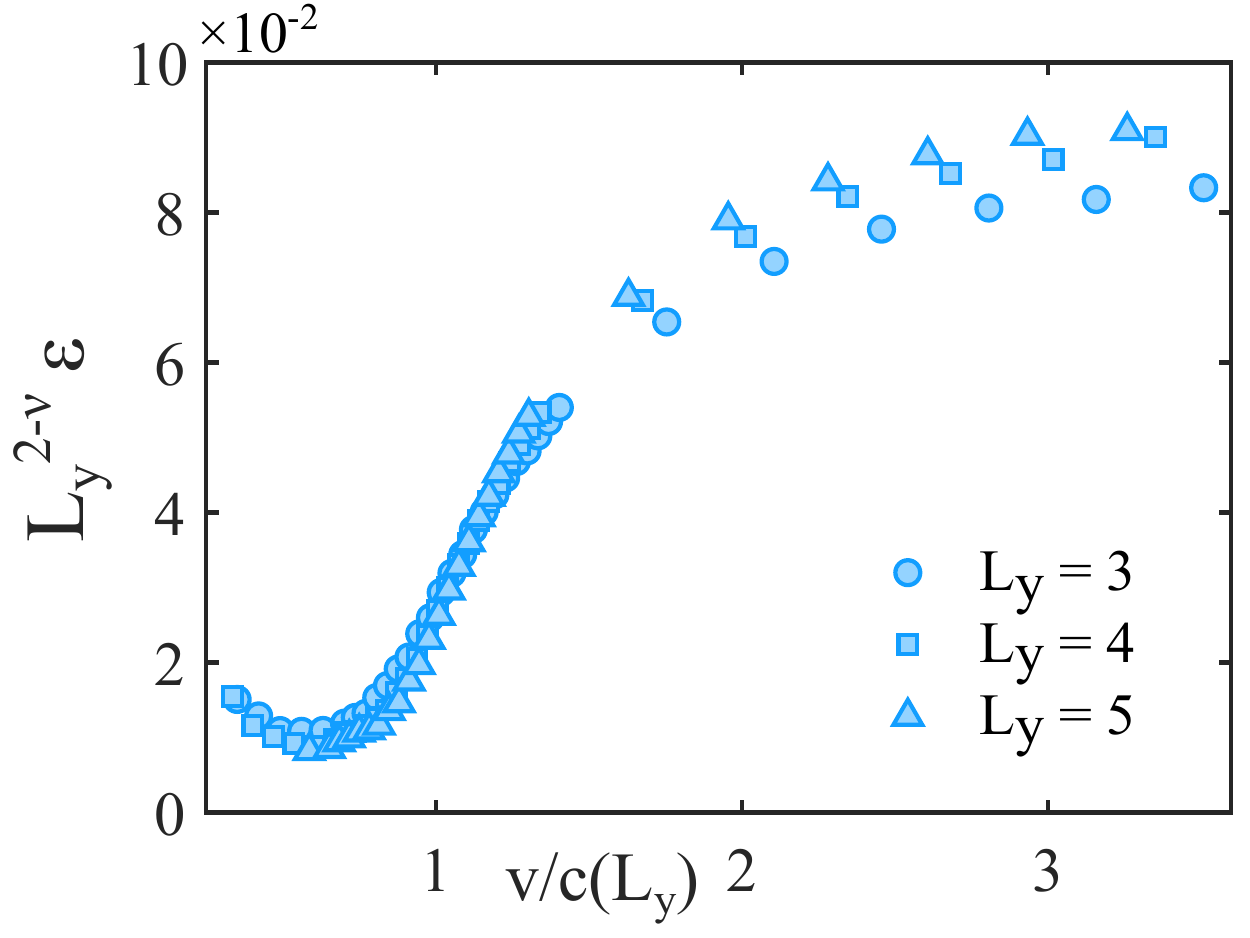}
    \caption{The total energy density is scaled with $L_y^{2-\nu}$ at $\tau=0.4$ and plotted on a linear scale as a function of $v/c(L_y)$. It shows poor scaling for $v/c(L_y)\geq1.5$ and is added to complement Fig.~\ref{fig:energyScaling} of the main text.}
    \label{fig:energyScalingLinear}
\end{figure}

In the main text of this paper, the finite size scaling of the energy density for quench velocities near the emergent speed of sound in each system is shown in Fig.~\ref{fig:critEnTotal}. However, the total post-quench energy density does not show a good scaling collapse for every quench velocities. It is clearer when the energy is plotted on a linear scale, as shown in Fig.~\ref{fig:energyScalingLinear}. Instead, it indicates that the system is excited for $v/c(L_y)\geq 1.5$ and fails to scale as the equilibrium critical theory. The region $0.3\leq v/c(L_y)\leq 1.4$ is where the data collapses well with $L_y^{2-\nu}$ as a function of $v/c(L_y)$.

\bibliography{manuscript}

\end{document}